%% file: jaipur.tex
\documentstyle[12pt,psfig]{article}
\newcommand{\be}{\begin{equation}}
\newcommand{\ee}{\end{equation}}
\newcommand{\bea}{\begin{array}{c}}
\newcommand{\eaa}{\end{array}}
\newcommand{\ba}{\begin{eqnarray}}
\newcommand{\ea}{\end{eqnarray}}

\date{\today}
\title{Strangeness :\\
Theoretical Status\footnote{Talk presented at the ``Third International
Conference on Physics and Astrophysics of Quark-Gluon Plasma'', March 17-21,
Jaipur, India.}}
\author{J.~Cleymans
\\
Department of Physics, University of Cape Town, South Africa}
\begin{document}
\maketitle
\input{psbox}
%
%
\begin{abstract}
Strangeness plays an important role in the study of quark matter since 
it indicates the rate at which new particles are being produced and 
therefore provides information about the degree of chemical equilibration
reached in heavy ion collisions. 
It is shown that ratios 
provide an excellent method for 
determining  the temperature
and the chemical potential since many theoretical uncertainties
cancel out. All indications are that 
particle abundances are very close 
to chemical equilibrium. 
The amount of strangeness present in $S-S$ collisions 
is very close to the
one corresponding to chemical equilibrium. Deviations are at the
30\% level.
\end{abstract}
\section{Introduction}
Considerable efforts have been made  to determine
the abundance of strange and other particles
in ultra-relativistic heavy ion collisions
in order to establish the composition  of the final state and to
infer about the history of the produced system~\cite{qm96,NA44,NA49}.
Due  to  the  complexity of the final state it is natural to
attempt  a statistical description based on the hadronic gas
model. This has been done by several groups recently with
considerable
success~\cite{stachel1,stachel2,
becattini1,jc-de,becattini2,rafelski1,sollfrank,apostolos}.

Since
the hadronic gas model uses statistical weight factors there are  only
very few parameters to describe the freeze-out stage of the produced
system.
In  the  Boltzmann approximation the
number of particles of type $i$, $N_i^0$, is
given by
\begin{eqnarray}
N_i^0 &=& g_iV\int {d^3p\over (2\pi)^3} e^{-E/T+\mu_i/T} \nonumber \\
      &=& {g_im_i^2T\over 2\pi^2}K_2(m_i/T)e^{\mu_i/T}
\end{eqnarray}
where $g_i$ is the degeneracy factor and  the index i refers to the
particle species under consideration.
The  corresponding  particle density  will  be  denoted  by $n_i^0\equiv
N_i^0/V$.

In  the hadron gas   model~\cite{jc-hs}  one  assumes  that  all  hadronic
resonances  contribute  according  to  their statistical weight
factors. At freeze-out all hadronic resonances decay into the lightest
stable particles.
All these
decays are taken into account  by using the known branching ratios.
\begin{equation}
N_{\pi^+}=\sum_iN_i^0 Br(i\rightarrow \pi^+)
\end{equation}
Figure  1  gives  an  indication of the importance
of resonances.
In this figure we plot the ratio of thermal pions
over the number of all pions
(i.e. including those coming from resonance decays).
If the temperature is low, resonances are unimportant and the
ratio is approximately one.
However at a temperature of $T =$ 200 MeV this ratio is
down to the  10\%  level.
Thus to a good approximation \underbar{there are no
thermal pions}\underbar{ if the}\underbar{ temperature}
\underbar{is above 200 MeV}.

For a realistic comparison with experimental data
it will be necessary to modify the above expression
in  several ways:
to  name  but a few, the fireball will not be produced at rest,
there  will  be  hard  core  interactions,  there  could  be  a
superposition  of  fireballs, flow effects~\cite{NA44,NA49}
etc...  The  expression for the particle
number given above will thus have to be modified.
We  will  show  below that many of these effects however cancel
out  when  considering  hadronic  ratios.  Thus if $N_i$ is the
particle  number  taking  into account various effects, then in
many cases one will have the remarkable result
\begin{equation}
{N_i\over N_j} = {N_i^0\over N_j^0}
\end{equation}
where $N_i$ refers to the particle abundance inclusive of effects due
to flow, transverse expansion, excluded volume while $N_i^0$ is the
particle abundance calculated in a straightforward manner using
Boltzmann statistics without modification, as given above.
This is of course not always the case but for several interesting
cases which have been widely discussed it is.
Therefore
\underbar{hadronic}\underbar{ ratios can be used}
\underbar{to determine basic}
\underbar{thermodynamic}
\underbar{quantities}
\underbar{like the}
\underbar{freeze-out}
\underbar{temperature}
\underbar{and}
\underbar{baryon}
\underbar{chemical}
\underbar{potential}
in  a  way  which  is  free  from  many  distortions (e.g.
transverse flow).
\begin{figure}
\begin{center}\mbox{\psboxscaled{500}{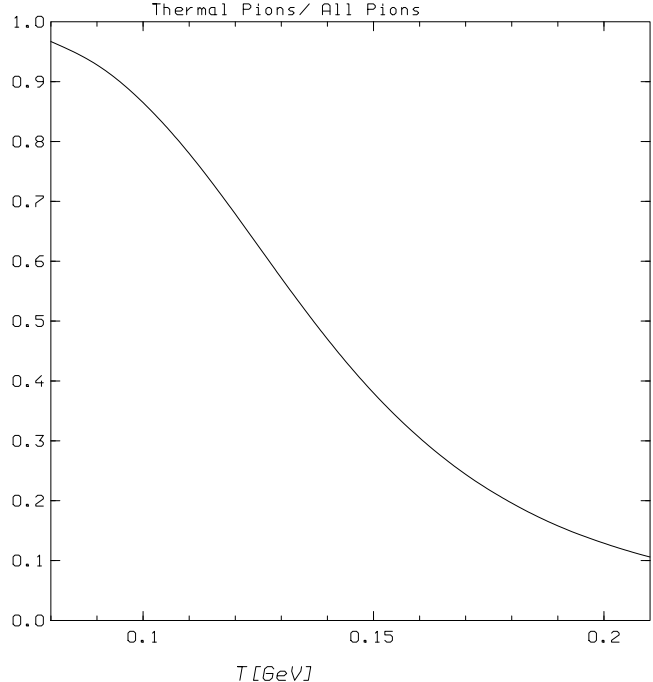}}\end{center}
\vskip 0.2cm
\caption{{\it Ratio  of  thermal pions to all pions (including pions
from resonance decays).}}
\end{figure}
We will consider in particular the following :
1) excluded volume corrections cancel out if the particle ratios involve an
integration over $4\pi$,
2) the effects due to a 
superposition of fireballs    cancel out for ratios provided the
abundances are integrated over all values of rapidity,
3) the effects of flow arising from instantaneous freeze-out 
cancel out if one considers ratios
integrated over transverse momenta,
4) for Bjorken longitudinal expansion accompanied by transverse
flow, the effects cancel out if one considers particle numbers integrated
over all transverse momenta.

Before  discussing this in more detail we  present
quantitatively what is  meant by chemical
equilibrium.
\section{Chemical Equilibrium}
Hadronic matter has a certain composition of particle species,
characterized by ``chemical'' potentials.
In addition
to the baryon number chemical potential, $\mu_B$, we have
a strangeness chemical potential, $\mu_S$, and also a charge chemical
potential $\mu_Q$..
The value of $\mu_B$ is fixed by giving the overall baryon number
density $n_B$, and that of $\mu_S$ by fixing the overall strangeness
to zero. The value of $\mu_Q$ is fixed by the total baryon number $B$
and the total charge $Q$. If $B/2Q$=1 the system is symmetric in
isospin and the charge chemical potential will be zero : $\mu_Q=0$.
This is the case for $S-S$ scattering but not for $p-p$ and also
not for $Pb-Pb$ where $B/2Q=0$ and  1.5 respectively.
For given values of $T$, $\mu_B$
and $\mu_Q$ one tunes the remaining parameter, $\mu_S$,
in such a way as to ensure strangeness
neutrality.
An example is given in table 1 where
we show the relevant Boltzmann factors for
different particles in the case of  chemical equilibrium.
If there is no chemical equilibrium the particle abundances are
unrestricted and their normalizations are arbitrary. Dynamical models are
then necessary  to establish particle abundances~\cite{knoll,spieles}.
If the system is not in chemical equilibrium than there is no relation
between different chemical potentials. In fact the number of chemical
potentials necessary to describe the state of the system gives an idea of
the degree of equilibration taking place.
As  perfect  chemical  equilibrium is of course an extreme case
one  would  like  to  parametrize deviations from in a simple
way. It has been known since a long time that there is a strangeness
deficiency in $p-p$ collisions,
A qualitative analysis of this was done
by Wroblewski~\cite{wroblewski} many years ago.
The deviation from equilibrium can be parameterized by a quantity
$\gamma_S$~\cite{rafelski,slotta}. Recent estimates~\cite{becattini2} give
$\gamma_S\approx 0.5$ in $p-p$ collisions. In relativistic heavy ion
collisions all estimates lead to a larger value $\gamma_S\approx 0.7 - 1.0$.

We believe that it is very important to establish
whether or not
chemical equilibrium has been reached.

\vskip 1 truecm
\centerline{
$$\vbox{\offinterlineskip \tabskip=0pt
{\offinterlineskip \tabskip=0pt
\halign{ \strut \vrule#&
\quad # \quad &      
      \vrule#&          
      \quad             
      \hfil # \quad &   %
      \vrule#&          
      \quad             
      \hfil # \quad &   %
      \vrule#&          
      \quad             
      \hfil # \quad &   %
      \vrule#          
      \cr               %
\noalign{\hrule}
&                           &&                         &&              & \cr
&       &&Chemical Equilibrium\hfil       && No Chemical Equilibrium& \cr
&                           &&                         &&              & \cr
\noalign{\hrule}                
&                           &&                         &&              & \cr
&$\pi$ &&$\displaystyle \exp\left[-{E_\pi\over T}\right]$
       &&$\displaystyle \exp\left[ -{E_\pi\over T}+{\mu_\pi\over T}\right]$ &\cr
&                           &&                         &&              & \cr
&$N$ &&$\displaystyle \exp\left[ -{E_N\over T}+ {\mu_B\over T}\right]$ 
     &&$\displaystyle \exp\left[ -{E_N\over T}+{\mu_N\over T}\right]$ &\cr
&                           &&                         &&              & \cr
&$\overline{N}$ &&$\displaystyle \exp\left[ -{E_N\over T} - {\mu_B\over T}\right]$ 
     &&$\displaystyle \exp\left[ -{E_N\over T}+{\mu_{\overline{N}}\over T}\right]$ &\cr
&                           &&                         &&              & \cr
&$\Lambda$ 
&&$\displaystyle \exp\left[ -{E_\Lambda\over T}+{\mu_B\over T}-{\mu_S\over T}\right]$ &&
$\displaystyle \exp\left[ -{E_\Lambda\over T}+{\mu_\Lambda\over T}\right]$ &\cr
&                           &&                         &&              & \cr
&$\overline{\Lambda}$
&&$\displaystyle \exp\left[ -{E_\Lambda\over T}-{\mu_B\over T}+{\mu_S\over T}\right]$ &&
$\displaystyle \exp\left[ -{E_\Lambda\over T}+{\mu_{\overline{\Lambda}}\over T}\right]$ &\cr
&                           &&                         &&              & \cr
&$K$ &&$\displaystyle \exp\left[ -{E_K\over T}+{\mu_S\over T}\right]$ 
     &&$\displaystyle \exp\left[ -{E_K\over T}+{\mu_K\over T}\right]$ &\cr
&                           &&                         &&              & \cr
&$\overline{K}$ &&$\displaystyle \exp\left[ -{E_K\over T}-{\mu_S\over T}\right]$ 
     &&$\displaystyle \exp\left[ -{E_K\over T}+{\mu_{\overline{K}}\over T}\right]$ &\cr
&                           &&                         &&              & \cr
\noalign{\hrule}}
}
}$$
}
\vskip 0.5 truecm
{\it Table 1 : Boltzmann factors for a hadronic gas
with B/2Q = 1 in thermal equilibrium.}
\vskip 1 truecm

\section{Hadronic Ratios.}
\subsection{Excluded Volume Corrections.}
In  many  cases  one  includes  the hard core repulsive nuclear
force in the
statistical description by giving hadrons a proper volume.
In many  cases~\cite{crss1,gorenstein} the particle 
density is modified in a geometric
way leading to the following form
\begin{equation}
\displaystyle
{n_i\over    n_j}   =
 {{n_i^0\over 1 + \sum_l n_l^0V_0}
   \over  {n_j^0\over 1 + \sum_l n_l^0V_0}}
\end{equation}
In this case it is clear that the corrections due to the hard core
repulsion cancel out  in the ratio of different hadron species and
one obtains
\begin{equation}
{n_i\over  n_j}   =  {n_i^0\over  n_j^0}
\end{equation}
This also holds for the excluded volume corrections considered in
ref.~\cite{gorenstein} 
(provided one uses the Boltzmann approximation) 
but not for the one considered in ref.~\cite{stachel1}.
\subsection{Superposition of Fireballs.}
Thermal models based on a single fireball produced in the
center-of-mass of the collisions fail to reproduce the observed rapidity
distribution. To remedy this one often considers that a 
distribution of
fireballs is being produced. For a symmetric system like $S-S$ or
$Pb-Pb$ it is  a good start to consider a uniform
distribution of fireballs along the rapidity axis but it
could be more generally described by a function
$\rho(Y_{FB})$ where $Y_{FB}$ is the rapidity of the fireball under
consideration. The particle abundance resulting from such a
distribution is obtained by integrating over all possible sources.
The ratio of two hadronic species is thus given by
\begin{equation}
{n_i\over    n_j}   =
 {\int_{-\infty}^{\infty}dy
          \int_{-Y}^YdY_{FB}~\rho(Y_{FB})
           {dn_i^0\over dy}(y-Y_{FB})
   \over \int_{-\infty}^{\infty}dy
          \int_{-Y}^Y dY_{FB}~\rho(Y_{FB})
          { dn_j^0\over dy}(y-Y_{FB})}
\end{equation}
If all the fireballs have the same temperature and the same chemical
potential then, after changing the order of integration, one obtains
\begin{equation}
{n_i\over n_j}~~~=~~~{n_i^0\over n_j^0}~{\int_{-Y}^Y~dY_{FB}~\rho (Y_{FB})
   \over\int_{-Y}^Y~dY_{FB}~\rho (Y_{FB})}
\end{equation}
as the integral over fireball distributions is the same in both
cases, one obtains again
\begin{equation}
\displaystyle {n_i\over  n_j}   =  {n_i^0\over  n_j^0}
\end{equation}
and the  effects   cancel out.
\subsection{Transverse Flow}
Recent results from $Pb-Pb$ collisions show that heavy particles like
protons and deuterons have a much larger transverse momentum than
light particles such as pions and kaons~\cite{qm96,NA44,NA49}.
Within the framework of thermal
models this indicates that fireballs have flow in the transverse
or at the very least have a random walk type of distribution in the
transverse velocity plane.
The hydrodynamic description has been extensively discussed in
recent papers~\cite{crs1,ruuskanen}.
The transverse momentum dependence can also be explained by assuming
fireballs which have  a gaussian distribution in the transverse
rapidity.
This can be due to either hydrodynamic expansion with flow
in the transverse direction or to fireballs being produced with a
random walk type of distribution in transverse
rapidity~\cite{leonidov,esumi}.
For the discussion about transverse flow
it is helpful to have the following
two analytic results
\begin{equation}
\int_0^\infty       dp_T~p_T
m_T^i~~K_1\left({m_T^i\over T}\cosh y_T\right)
I_0\left({ p_T\over T}\sinh y_T\right)
=m_i^2TK_2\left({m^i\over T}\right)\cosh y_T
\end{equation}
and
\begin{equation}
\int_0^\infty       dp_T~p_T
p_T~K_0\left({m_T^i\over T}\cosh y_T\right)
I_1\left({ p_T\over T}\sinh y_T\right)
=m_i^2TK_2\left({m^i\over T}\right)\sinh y_T
\end{equation}
\subsubsection{Instantaneous Freeze-out}
If the system undergoes instantaneous freeze-out, the mathematical
treatment simplifies considerably since no
assumptions have to be made about the
freeze-out surface. The velocity profile at
freeze-out time can be arbitrary but the system should have the same freeze-out
temperature everywhere.
The expression for the particle abundance is given by :
\begin{equation}
{\displaystyle
n_i\over    n_j}   =
 {\int r~dr\int_0^\infty       dp_T
m_T^i~~K_1\left({\displaystyle m_T^i\over T}\cosh y_T\right)
I_0\left({\displaystyle p_T\over T}\sinh y_T\right)
\over \int r~dr
\int_0^\infty       dp_T
m_T^j~~K_1\left({\displaystyle m_T^j\over T}\cosh y_T\right)
I_0\left({\displaystyle p_T\over T}\sinh y_T\right)}
\end{equation}
Using the integrals given above in eqs. (9) and (10), we obtain
\begin{equation}
{\displaystyle
n_i\over  n_j}   =  {m_i^2~~K_2\left({\displaystyle m^i\over T}\right)
\int r~dr~\cosh y_T
\over
m_j^2~~K_2\left({\displaystyle m^i\over T}\right)\int r~dr~\cosh y_T}
\end{equation}
so that all effects of flow cancel out in the ratio and we are left
with the result
\begin{equation}
{n_i\over  n_j}   =  {n_i^0\over  n_j^0} .
\end{equation}
A very similar cancellation also occurs in
other models, e.g. the model that was recently
considered in ref.~\cite{leonidov} based on
fireballs having a random walk
type of distribution  in the transverse
direction leads to the same  cancellation in the hadronic ratios.
\subsection{ Bjorken scaling + Transverse expansion}
If freeze-out is not instantaneous an extra term appears in the expression for
the particle spectrum~\cite{crs1,ruuskanen}
\begin{eqnarray}
\left( {dN_i\over dy m_Tdm_T}\right)_{y=0}
&  &   = {g\over \pi} \int_\sigma r~dr~\tau_F(r)    \nonumber\\
& &\left\{ m_T^iI_0 \left( {p_T\sinh y_T\over T} \right)\right.
          K_1 \left( {m_T^i\cosh y_T\over T} \right) \nonumber \\
-& &\left( {\partial\tau_F\over\partial r} \right) p_T
          I_1 \left( {p_T\sinh y_T\over T} \right)
    \left. K_0 \left( {m_T^i\cosh y_T\over T} \right) \right\}
\end{eqnarray}
where the partial derivative $\partial\tau_F/\partial r $ has to be taken
on the freeze-out surface, denoted by $\sigma$ in the equation above.
After integration over $m_T$ (and only after this integration) one obtains
\begin{eqnarray}
\left( {dN_i\over dy }\right)_{y=0}
= & &{g\over \pi} \int_\sigma r~dr~\tau_F(r)    \nonumber \\
& & \left\{ \cosh(y_T) - \left({\partial\tau_F\over\partial r} \right)
\sinh (y_T) \right\} m_i^2 T K_2 \left( {m_i\over T} \right)
\end{eqnarray}
leading to the result
\begin{equation}
{dN_i/dy\over dN_j/dy} = {N_i^0\over N_j^0}
\end{equation}
This result is remarkable : in a model where the transverse spectra
are severely distorted by transverse flow and the rapidity
distribution has a plateau, the particle ratios are still given by
the uncorrected Boltzmann distribution as if particles originated
from a fireball at rest.
Effects of hydrodynamic flow cancel out in the ratio provided
the particle abundance has been integrated over transverse momentum.
\\
\section{Results}
The above discussion has given us some confidence that the particle
ratios can be used to determine the temperature and the chemical
potential in a given reaction. 
Comparison of hadronic abundances with thermal models has
 been done recently for $p-p$ collisions~\cite{becattini2}, 
 for $Si-Au$ \cite{stachel1,jc-de} and
 $Au-Au$ \cite{stachel2} collisions at BNL, and also 
 for the preliminary results
 of the results from $Pb-Pb$ at CERN.
%
Each particle ratio leads to a band in the $T, \mu_B$ plane,
(including the error bar) this gives rise to a region where the
hadronic gas reproduces the measured numbers. If all the measured
ratios have a common overlap region then one can conclude that the
experimental data are consistent with chemical equilibrium.
\begin{figure}
\begin{center}\mbox{\psboxscaled{500}{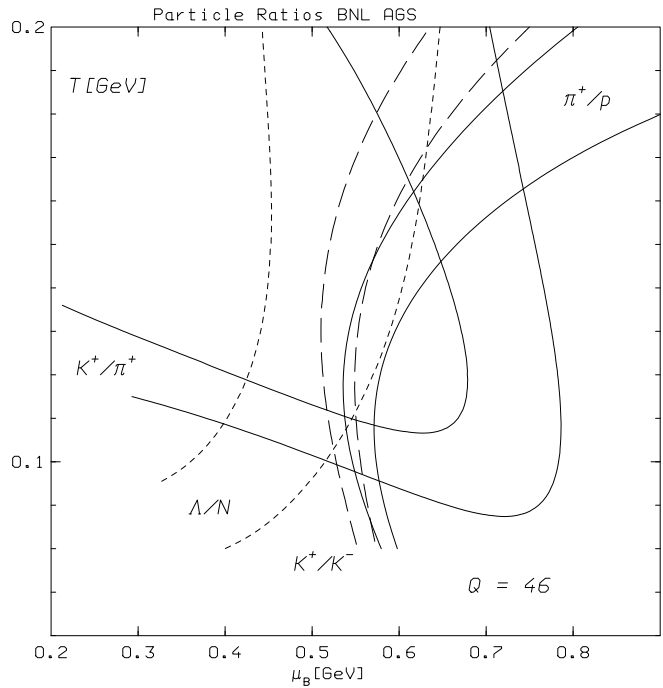}}\end{center}
\vskip 0.2cm
\caption{{\it  Particle ratios using data from the AGS at BNL.}}
\end{figure}
The data from Brookhaven lead to $T\approx 100-120$ Mev and a chemical
potential which is surprisingly well determined $\mu_B$=540 MeV.
The CERN data give $T\approx 180$ MeV and a chemical potential 
which is substantially lower than the one from BNL $\mu_B\approx$
200-300 MeV.
\begin{figure}
\begin{center}\mbox{\psboxscaled{500}{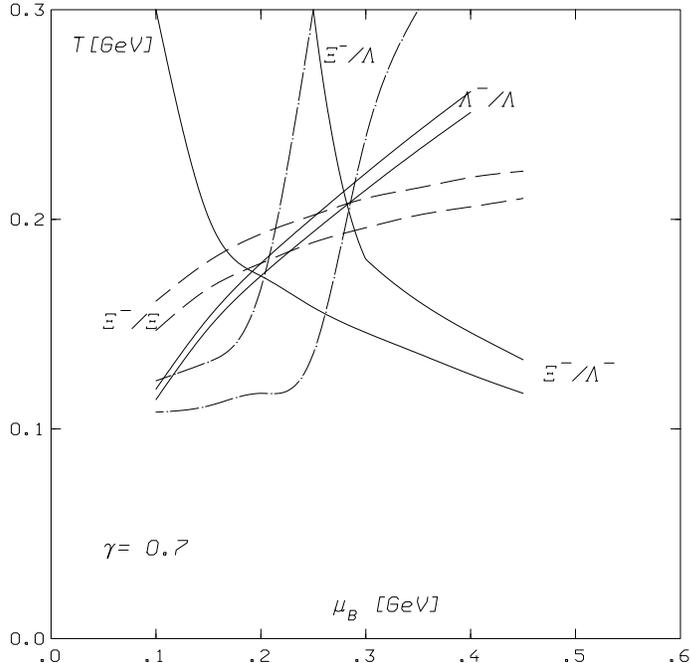}}\end{center}
\vskip 0.2cm
\caption{{\it  Strange particle ratios using data from the WA85
collaboration.}}
\end{figure}
A discussion of different break-up mechanisms has been presented in
\cite{redlich,heinz,crss}.\\
\section{Conclusions}
One of the important parameters in the description of relativistic
heavy ion collisions is the temperature. We have argued that
the best way to determine
directly the temperature is to consider ratios of hadronic yields
that have been integrated over transverse momentum. In such a ratio
many effects (e.g. flow) cancel out. A surprisingly large number of
hadronic abundances can be explained in this way with a temperature
which is around 180 MeV for data obtained from $S-S$ collisions 
at CERN and
around 110 MeV for data from the AGS.
One conclusion forces itself upon us : there is nothing
strange about strangeness.
\\
\subsection*{Acknowledgment}
\hspace*{\parindent}
We gratefully acknowledge  useful and stimulating
discussions with Helmut Satz, Krzysztof Redlich and Dinesh Srivastava.
The  financial support of the University of Cape Town (URC) and
of the Foundation for Research Development (FRD) are gratefully
acknowledged.
\end{document}

%% file: psbox.tex
\def\temp{1.34}%
\let\tempp=\relax
\expandafter\ifx\csname psboxversion\endcsname\relax
  \message{PSBOX(\temp) loading}%
\else
    \ifdim\temp cm>\psboxversion cm
      \message{PSBOX(\temp) loading}%
    \else
      \message{PSBOX(\psboxversion) is already loaded: I won't load
        PSBOX(\temp)!}%
      \let\temp=\psboxversion
      \let\tempp= 
    \fi
\fi
\tempp
\let\psboxversion=\temp
\catcode`\@=11
%
%
\def\psfortextures{
\def\PSspeci@l##1##2{%
\special{illustration ##1\space scaled ##2}%
}}%
\def\psfordvitops{
\def\PSspeci@l##1##2{%
\special{dvitops: import ##1\space \the\drawingwd \the\drawinght}%
}}%
\def\psfordvips{
\def\PSspeci@l##1##2{%
\d@my=0.1bp \d@mx=\drawingwd \divide\d@mx by\d@my
\includegraphics{##1\space}}}%
\def\psforoztex{
\def\PSspeci@l##1##2{%
\special{##1 \space
      ##2 1000 div dup scale
      \number-\psllx\space \number-\pslly\space translate
}}}%
\def\psfordvitps{
\def\psdimt@n@sp##1{\d@mx=##1\relax\edef\psn@sp{\number\d@mx}}
\def\PSspeci@l##1##2{%
\special{dvitps: Include0 "psfig.psr"}
\psdimt@n@sp{\drawingwd}
\special{dvitps: Literal "\psn@sp\space"}
\psdimt@n@sp{\drawinght}
\special{dvitps: Literal "\psn@sp\space"}
\psdimt@n@sp{\psllx bp}
\special{dvitps: Literal "\psn@sp\space"}
\psdimt@n@sp{\pslly bp}
\special{dvitps: Literal "\psn@sp\space"}
\psdimt@n@sp{\psurx bp}
\special{dvitps: Literal "\psn@sp\space"}
\psdimt@n@sp{\psury bp}
\special{dvitps: Literal "\psn@sp\space startTexFig\space"}
\special{dvitps: Include1 "##1"}
\special{dvitps: Literal "endTexFig\space"}
}}%
\def\psfordvialw{
\def\PSspeci@l##1##2{
\special{language "PostScript",
position = "bottom left",
literal "  \psllx\space \pslly\space translate
  ##2 1000 div dup scale
  -\psllx\space -\pslly\space translate",
include "##1"}
}}%
\def\psforptips{
\def\PSspeci@l##1##2{{
\d@mx=\psurx bp
\advance \d@mx by -\psllx bp
\divide \d@mx by 1000\multiply\d@mx by \xscale
\incm{\d@mx}
\let\tmpx\dimincm
\d@my=\psury bp
\advance \d@my by -\pslly bp
\divide \d@my by 1000\multiply\d@my by \xscale
\incm{\d@my}
\let\tmpy\dimincm
\d@mx=-\psllx bp
\divide \d@mx by 1000\multiply\d@mx by \xscale
\d@my=-\pslly bp
\divide \d@my by 1000\multiply\d@my by \xscale
\at(\d@mx;\d@my){\special{ps:##1 x=\tmpx, y=\tmpy}}
}}}%
\def\psonlyboxes{
\def\PSspeci@l##1##2{%
\at(0cm;0cm){\boxit{\vbox to\drawinght
  {\vss\hbox to\drawingwd{\at(0cm;0cm){\hbox{({\tt##1})}}\hss}}}}
}}%
\def\psloc@lerr#1{%
\let\savedPSspeci@l=\PSspeci@l%
\def\PSspeci@l##1##2{%
\at(0cm;0cm){\boxit{\vbox to\drawinght
  {\vss\hbox to\drawingwd{\at(0cm;0cm){\hbox{({\tt##1}) #1}}\hss}}}}
\let\PSspeci@l=\savedPSspeci@l
}}%
%
%
\newread\pst@mpin
\newdimen\drawinght\newdimen\drawingwd
\newdimen\psxoffset\newdimen\psyoffset
\newbox\drawingBox
\newcount\xscale \newcount\yscale \newdimen\pscm\pscm=1cm
\newdimen\d@mx \newdimen\d@my
\newdimen\pswdincr \newdimen\pshtincr
\let\ps@nnotation=\relax
{\catcode`\|=0 |catcode`|\=12 |catcode`|
|catcode`#=12 |catcode`*=14
|xdef|backslashother{\}*
|xdef|percentother{
|xdef|tildeother{~}*
|xdef|sharpother{#}*
}%
\def\R@moveMeaningHeader#1:->{}%
\def\uncatcode#1{%
\edef#1{\expandafter\R@moveMeaningHeader\meaning#1}}%
\def\execute#1{#1}
\def\psm@keother#1{\catcode`#112\relax}
\def\executeinspecs#1{%
\execute{\begingroup\let\do\psm@keother\dospecials\catcode`\^^M=9#1\endgroup}}%
\def\@mpty{}%
\def\matchexpin#1#2{
  \fi%
  \edef\tmpb{{#2}}%
  \expandafter\makem@tchtmp\tmpb%
  \edef\tmpa{#1}\edef\tmpb{#2}%
  \expandafter\expandafter\expandafter\m@tchtmp\expandafter\tmpa\tmpb\endm@tch%
  \if\match%
}%
\def\matchin#1#2{%
  \fi%
  \makem@tchtmp{#2}%
  \m@tchtmp#1#2\endm@tch%
  \if\match%
}%
\def\makem@tchtmp#1{\def\m@tchtmp##1#1##2\endm@tch{%
  \def\tmpa{##1}\def\tmpb{##2}\let\m@tchtmp=\relax%
  \ifx\tmpb\@mpty\def\match{YN}%
  \else\def\match{YY}\fi%
}}%
\def\incm#1{{\psxoffset=1cm\d@my=#1
 \d@mx=\d@my
  \divide\d@mx by \psxoffset
  \xdef\dimincm{\number\d@mx.}
  \advance\d@my by -\number\d@mx cm
  \multiply\d@my by 100
 \d@mx=\d@my
  \divide\d@mx by \psxoffset
  \edef\dimincm{\dimincm\number\d@mx}
  \advance\d@my by -\number\d@mx cm
  \multiply\d@my by 100
 \d@mx=\d@my
  \divide\d@mx by \psxoffset
  \xdef\dimincm{\dimincm\number\d@mx}
}}%
%
\newif\ifNotB@undingBox
\newhelp\PShelp{Proceed: you'll have a 5cm square blank box instead of
your graphics (Jean Orloff).}%
\def\s@tsize#1 #2 #3 #4\@ndsize{
  \def\psllx{#1}\def\pslly{#2}%
  \def\psurx{#3}\def\psury{#4}
  \ifx\psurx\@mpty\NotB@undingBoxtrue
  \else
    \drawinght=#4bp\advance\drawinght by-#2bp
    \drawingwd=#3bp\advance\drawingwd by-#1bp
  \fi
  }%
\def\sc@nBBline#1:#2\@ndBBline{\edef\p@rameter{#1}\edef\v@lue{#2}}%
\def\g@bblefirstblank#1#2:{\ifx#1 \else#1\fi#2}%
{\catcode`\%=12
\xdef\B@undingBox{
\def\ReadPSize#1{
 \readfilename#1\relax
 \let\PSfilename=\lastreadfilename
 \openin\pst@mpin=#1\relax
 \ifeof\pst@mpin \errhelp=\PShelp
   \errmessage{I haven't found your postscript file (\PSfilename)}%
   \psloc@lerr{was not found}%
   \s@tsize 0 0 142 142\@ndsize
   \closein\pst@mpin
 \else
   \if\matchexpin{\GlobalInputList}{, \lastreadfilename}%
   \else\xdef\GlobalInputList{\GlobalInputList, \lastreadfilename}%
     \immediate\write\psbj@inaux{\lastreadfilename,}%
   \fi%
   \loop
     \executeinspecs{\catcode`\ =10\global\read\pst@mpin to\n@xtline}%
     \ifeof\pst@mpin
       \errhelp=\PShelp
       \errmessage{(\PSfilename) is not an Encapsulated PostScript File:
           I could not find any \B@undingBox: line.}%
       \edef\v@lue{0 0 142 142:}%
       \psloc@lerr{is not an EPSFile}%
       \NotB@undingBoxfalse
     \else
       \expandafter\sc@nBBline\n@xtline:\@ndBBline
       \ifx\p@rameter\B@undingBox\NotB@undingBoxfalse
         \edef\t@mp{%
           \expandafter\g@bblefirstblank\v@lue\space\space\space}%
         \expandafter\s@tsize\t@mp\@ndsize
       \else\NotB@undingBoxtrue
       \fi
     \fi
   \ifNotB@undingBox\repeat
   \closein\pst@mpin
 \fi
\message{#1}%
}%
%
%
\def\psboxto(#1;#2)#3{\vbox{
   \ReadPSize{#3}%
   \divide\drawingwd by 1000
   \divide\drawinght by 1000
   \d@mx=#1
   \ifdim\d@mx=0pt\xscale=1000
         \else \xscale=\d@mx \divide \xscale by \drawingwd\fi
   \d@my=#2
   \ifdim\d@my=0pt\yscale=1000
         \else \yscale=\d@my \divide \yscale by \drawinght\fi
   \ifnum\yscale=1000
         \else\ifnum\xscale=1000\xscale=\yscale
                    \else\ifnum\yscale<\xscale\xscale=\yscale\fi
              \fi
   \fi
   \divide\pswdincr by 1000 \multiply\pswdincr by \xscale
   \divide\pshtincr by 1000 \multiply\pshtincr by \xscale
   \divide\psxoffset by1000 \multiply\psxoffset by\xscale
   \divide\psyoffset by1000 \multiply\psyoffset by\xscale
   \global\divide\pscm by 1000
   \global\multiply\pscm by\xscale
   \multiply\drawingwd by\xscale \multiply\drawinght by\xscale
   \ifdim\d@mx=0pt\d@mx=\drawingwd\fi
   \ifdim\d@my=0pt\d@my=\drawinght\fi
   \message{scaled \the\xscale}%
 \hbox to\d@mx{\hss\vbox to\d@my{\vss
   \global\setbox\drawingBox=\hbox to 0pt{\kern\psxoffset\vbox to 0pt{
      \kern-\psyoffset
      \PSspeci@l{\PSfilename}{\the\xscale}%
      \vss}\hss\ps@nnotation}%
   \advance\pswdincr by \drawingwd
   \advance\pshtincr by \drawinght
   \global\wd\drawingBox=\the\pswdincr
   \global\ht\drawingBox=\the\pshtincr
   \baselineskip=0pt
   \copy\drawingBox
 \vss}\hss}%
  \global\psxoffset=0pt
  \global\psyoffset=0pt
  \global\pswdincr=0pt
  \global\pshtincr=0pt 
  \global\pscm=1cm 
  \global\drawingwd=\drawingwd
  \global\drawinght=\drawinght
}}%
%
%
\def\psboxscaled#1#2{\vbox{
  \ReadPSize{#2}%
  \xscale=#1
  \message{scaled \the\xscale}%
  \advance\drawingwd by\pswdincr\advance\drawinght by\pshtincr
  \divide\pswdincr by 1000 \multiply\pswdincr by \xscale
  \divide\pshtincr by 1000 \multiply\pshtincr by \xscale
  \divide\psxoffset by1000 \multiply\psxoffset by\xscale
  \divide\psyoffset by1000 \multiply\psyoffset by\xscale
  \divide\drawingwd by1000 \multiply\drawingwd by\xscale
  \divide\drawinght by1000 \multiply\drawinght by\xscale
  \global\divide\pscm by 1000
  \global\multiply\pscm by\xscale
  \global\setbox\drawingBox=\hbox to 0pt{\kern\psxoffset\vbox to 0pt{
     \kern-\psyoffset
     \PSspeci@l{\PSfilename}{\the\xscale}%
     \vss}\hss\ps@nnotation}%
  \advance\pswdincr by \drawingwd
  \advance\pshtincr by \drawinght
  \global\wd\drawingBox=\the\pswdincr
  \global\ht\drawingBox=\the\pshtincr
  \baselineskip=0pt
  \copy\drawingBox
  \global\psxoffset=0pt
  \global\psyoffset=0pt
  \global\pswdincr=0pt
  \global\pshtincr=0pt 
  \global\pscm=1cm
  \global\drawingwd=\drawingwd
  \global\drawinght=\drawinght
}}%
%
\def\psbox#1{\psboxscaled{1000}{#1}}%
\newif\ifn@teof\n@teoftrue
\newif\ifc@ntrolline
\newif\ifmatch
\newread\j@insplitin
\newwrite\j@insplitout
\newwrite\psbj@inaux
\immediate\openout\psbj@inaux=psbjoin.aux
\immediate\write\psbj@inaux{\string\joinfiles}%
\immediate\write\psbj@inaux{\jobname,}%
%
%
\def\toother#1{\ifcat\relax#1\else\expandafter%
  \toother@ux\meaning#1\endtoother@ux\fi}%
\def\toother@ux#1 #2#3\endtoother@ux{\def\tmp{#3}%
  \ifx\tmp\@mpty\def\tmp{#2}\let\next=\relax%
  \else\def\next{\toother@ux#2#3\endtoother@ux}\fi%
\next}%
%
%
\let\readfilenamehook=\relax
\def\re@d{\expandafter\re@daux}
\def\re@daux{\futurelet\nextchar\stopre@dtest}%
\def\re@dnext{\xdef\lastreadfilename{\lastreadfilename\nextchar}%
  \afterassignment\re@d\let\nextchar}%
\def\stopre@d{\egroup\readfilenamehook}%
\def\stopre@dtest{%
  \ifcat\nextchar\relax\let\nextread\stopre@d
  \else
    \ifcat\nextchar\space\def\nextread{%
      \afterassignment\stopre@d\chardef\nextchar=`}%
    \else\let\nextread=\re@dnext
      \toother\nextchar
      \edef\nextchar{\tmp}%
    \fi
  \fi\nextread}%
\def\readfilename{\vbox\bgroup%
  \let\\=\backslashother \let\%=\percentother \let\~=\tildeother
  \let\#=\sharpother \xdef\lastreadfilename{}%
  \re@d}%
%
%
\xdef\GlobalInputList{\jobname}%
\def\psnewinput{%
  \def\readfilenamehook{
    \if\matchexpin{\GlobalInputList}{, \lastreadfilename}%
    \else\xdef\GlobalInputList{\GlobalInputList, \lastreadfilename}%
      \immediate\write\psbj@inaux{\lastreadfilename,}%
    \fi%
    \ps@ldinput\lastreadfilename\relax%
    \let\readfilenamehook=\relax%
  }\readfilename%
}%
\expandafter\ifx\csname @@input\endcsname\relax    
  \immediate\let\ps@ldinput=\input\def\input{\psnewinput}%
\else
  \immediate\let\ps@ldinput=\@@input
  \def\@@input{\psnewinput}%
\fi%
\def\nowarnopenout{%
 \def\warnopenout##1##2{%
   \readfilename##2\relax
   \message{\lastreadfilename}%
   \immediate\openout##1=\lastreadfilename\relax}}%
\def\warnopenout#1#2{%
 \readfilename#2\relax
 \def\t@mp{TrashMe,psbjoin.aux,psbjoint.tex,}\uncatcode\t@mp
 \if\matchexpin{\t@mp}{\lastreadfilename,}%
 \else
   \immediate\openin\pst@mpin=\lastreadfilename\relax
   \ifeof\pst@mpin
     \else
     \errhelp{If the content of this file is so precious to you, abort (ie
press x or e) and rename it before retrying.}%
     \errmessage{I'm just about to replace your file named \lastreadfilename}%
   \fi
   \immediate\closein\pst@mpin
 \fi
 \message{\lastreadfilename}%
 \immediate\openout#1=\lastreadfilename\relax}%
{\catcode`\%=12\catcode`\*=14
\gdef\splitfile#1{*
 \readfilename#1\relax
 \immediate\openin\j@insplitin=\lastreadfilename\relax
 \ifeof\j@insplitin
   \message{! I couldn't find and split \lastreadfilename!}*
 \else
   \immediate\openout\j@insplitout=TrashMe
   \message{< Splitting \lastreadfilename\space into}*
   \loop
     \ifeof\j@insplitin
       \immediate\closein\j@insplitin\n@teoffalse
     \else
       \n@teoftrue
       \executeinspecs{\global\read\j@insplitin to\spl@tinline\expandafter
         \ch@ckbeginnewfile\spl@tinline
       \ifc@ntrolline
       \else
         \toks0=\expandafter{\spl@tinline}*
         \immediate\write\j@insplitout{\the\toks0}*
       \fi
     \fi
   \ifn@teof\repeat
   \immediate\closeout\j@insplitout
 \fi\message{>}*
}*
\gdef\ch@ckbeginnewfile#1
 \def\t@mp{#1}*
 \ifx\@mpty\t@mp
   \def\t@mp{#3}*
   \ifx\@mpty\t@mp
     \global\c@ntrollinefalse
   \else
     \immediate\closeout\j@insplitout
     \warnopenout\j@insplitout{#2}*
     \global\c@ntrollinetrue
   \fi
 \else
   \global\c@ntrollinefalse
 \fi}*
\gdef\joinfiles#1\into#2{*
 \message{< Joining following files into}*
 \warnopenout\j@insplitout{#2}*
 \message{:}*
 {*
 \edef\w@##1{\immediate\write\j@insplitout{##1}}*
\w@{
\w@{
\w@{
\w@{
\w@{
\w@{
\w@{
\w@{
\w@{
\w@{
\w@{\string\input\space psbox.tex}*
\w@{\string\splitfile{\string\jobname}}*
\w@{\string\let\string\autojoin=\string\relax}*
}*
 \expandafter\tre@tfilelist#1, \endtre@t
 \immediate\closeout\j@insplitout
 \message{>}*
}*
\gdef\tre@tfilelist#1, #2\endtre@t{*
 \readfilename#1\relax
 \ifx\@mpty\lastreadfilename
 \else
   \immediate\openin\j@insplitin=\lastreadfilename\relax
   \ifeof\j@insplitin
     \errmessage{I couldn't find file \lastreadfilename}*
   \else
     \message{\lastreadfilename}*
     \immediate\write\j@insplitout{
     \executeinspecs{\global\read\j@insplitin to\oldj@ininline}*
     \loop
       \ifeof\j@insplitin\immediate\closein\j@insplitin\n@teoffalse
       \else\n@teoftrue
         \executeinspecs{\global\read\j@insplitin to\j@ininline}*
         \toks0=\expandafter{\oldj@ininline}*
         \let\oldj@ininline=\j@ininline
         \immediate\write\j@insplitout{\the\toks0}*
       \fi
     \ifn@teof
     \repeat
   \immediate\closein\j@insplitin
   \fi
   \tre@tfilelist#2, \endtre@t
 \fi}*
}%
\def\autojoin{%
 \immediate\write\psbj@inaux{\string\into{psbjoint.tex}}%
 \immediate\closeout\psbj@inaux
 \expandafter\joinfiles\GlobalInputList\into{psbjoint.tex}%
}%
%
%
%
\def\centinsert#1{\midinsert\line{\hss#1\hss}\endinsert}%
\def\psannotate#1#2{\vbox{%
  \def\ps@nnotation{#2\global\let\ps@nnotation=\relax}#1}}%
\def\pscaption#1#2{\vbox{%
   \setbox\drawingBox=#1
   \copy\drawingBox
   \vskip\baselineskip
   \vbox{\hsize=\wd\drawingBox\setbox0=\hbox{#2}%
     \ifdim\wd0>\hsize
       \noindent\unhbox0\tolerance=5000
    \else\centerline{\box0}%
    \fi
}}}%
%
\def\at(#1;#2)#3{\setbox0=\hbox{#3}\ht0=0pt\dp0=0pt
  \rlap{\kern#1\vbox to0pt{\kern-#2\box0\vss}}}%
%
\newdimen\gridht \newdimen\gridwd
\def\gridfill(#1;#2){%
  \setbox0=\hbox to 1\pscm
  {\vrule height1\pscm width.4pt\leaders\hrule\hfill}%
  \gridht=#1
  \divide\gridht by \ht0
  \multiply\gridht by \ht0
  \gridwd=#2
  \divide\gridwd by \wd0
  \multiply\gridwd by \wd0
  \advance \gridwd by \wd0
  \vbox to \gridht{\leaders\hbox to\gridwd{\leaders\box0\hfill}\vfill}}%
%
\def\fillinggrid{\at(0cm;0cm){\vbox{%
  \gridfill(\drawinght;\drawingwd)}}}%
%
%
\def\textleftof#1:{%
  \setbox1=#1
  \setbox0=\vbox\bgroup
    \advance\hsize by -\wd1 \advance\hsize by -2em}%
\def\textrightof#1:{%
  \setbox0=#1
  \setbox1=\vbox\bgroup
    \advance\hsize by -\wd0 \advance\hsize by -2em}%
\def\endtext{%
  \egroup
  \hbox to \hsize{\valign{\vfil##\vfil\cr%
\box0\cr%
\noalign{\hss}\box1\cr}}}%
%
\def\frameit#1#2#3{\hbox{\vrule width#1\vbox{%
  \hrule height#1\vskip#2\hbox{\hskip#2\vbox{#3}\hskip#2}%
        \vskip#2\hrule height#1}\vrule width#1}}%
\def\boxit#1{\frameit{0.4pt}{0pt}{#1}}%
\catcode`\@=12 
%
 \psfordvips   

%% file: jaipur.bbl
\begin{thebibliography}{99}
\bibitem{qm96}
          {\it ``Quark Matter '96''},
          Proc. of the XII'th Int. Conf. on
          Ultra-Relativistic Nucleus-Nucleus Collisions, Heidelberg, 1996, 
P.~Braun-Munzinger, H.J.~Specht, R.~Stock and H.~St\"ocker (editors),
          Nucl. Phys. {\bf A610} (1996).
%
%
\bibitem{NA44} L.G. Bearden et al. (NA44 collaboration), Phys. Rev.
Letters {\bf 78}, 2080 (1997).
%
\bibitem{NA49} P. Jacobs at al. (NA49 collaboration) contribution to
this conference.
%
\bibitem{stachel1} P.  Braun-Munzinger,  J. Stachel, J. P. Wessels,
and N. Xu, Phys. Lett. {\bf B344}, 43 (1995); B {\bf 365}, 1 (1996).
%
%
%
\bibitem{stachel2} P.  Braun-Munzinger and  J. Stachel, 
Nucl. Phys. {\bf A606}, 320 (1996).
%
\bibitem{becattini1}  F. Becattini,
Z. f. Physik {\bf C69}, 485 (1996).
%
%
\bibitem{jc-de} J.~Cleymans, D.~Elliott, H.~Satz and R.L.~Thews,
Z. f. Physik {\bf C74}, 319 (1997).
%
%
\bibitem{becattini2}  F. Becattini and U.~Heinz,
Firenze preprint DFF 268/02/1997, 
Z. f. Physik {\bf C} to be published.
%
\bibitem{rafelski1} J. Letessier, A. Tounsi, U. Heinz, J. Sollfrank
and J. Rafelski, Phys. Rev. {\bf D51}, 3408 (1995).
%
\bibitem{sollfrank} J. Sollfrank et al., Z. f. Physik {\bf C61}, 659 (1994).
%
%
\bibitem{apostolos} M.N. Asprouli and A.D. Panagiotou, 
Phys. Rev. {\bf C51}, 1444 (1995).
%
%
\bibitem{jc-hs} See e.g. J.~Cleymans and H.~Satz, Z. f. Physik {\bf C57}, 135 (1993).
%
%
%
%
%
\bibitem{knoll} H.W. Barz, B.L. Friman, J. Knoll and H. Schulz, Nucl. Phys.
{\bf A484}, 661 (1988).
%
\bibitem{spieles} C. Spieles, H. St\"ocker and C.~Greiner, 
report nucl-th/9704008.
%
%
\bibitem{wroblewski} A. Wroblewski, Acta Phys. Pol. {\bf B16}, 379 (1985).
%
%
\bibitem{rafelski}  J.~Rafelski,
Nucl. Phys. {\bf A544}, 279c (1992).
%
\bibitem{slotta} C.~Slotta, J.~Sollfrank and U.~Heinz, in {\it ``Strangeness in
Hadronic Matter.''}, Ed. J. Rafelski, AIP Conference Proceedings 340, 1995.
%
%
\bibitem{crss1} J.~Cleymans, K.~Redlich, H.~Satz and E.~Suhonen,
Z. f. Physik {\bf C33}, 151 (1986).
%
%
\bibitem{gorenstein} D.K. Rischke,  M.~Gorenstein,  H. St\"ocker 
and W. Greiner, Z. f. Physik, {\bf C51}, 485 (1991).
%
%
\bibitem{crs1} J. Cleymans, K. Redlich, and D. K. Srivastava,
             Phys. Rev. {\bf C55}, 1431 (1997).
%
\bibitem{ruuskanen}J.  Sollfrank  et  al. Phys. Rev. {C55} 1431
(1997).
%
%
\bibitem{leonidov} A. Leonidov, M.Nardi and H. Satz,
                Z. f. Physik {\bf C74}, 535 (1997).
%
%
\bibitem{esumi} S. Esumi, U. Heinz and N. Xu, preprint  March 1997.
%
%
%
\bibitem{redlich}  K.~Redlich, J.~Cleymans, H.~Satz and E.~Suhonen,
Nucl. Phys. {\bf A566}, 391c (1994).
%
%
\bibitem{heinz} U.~Heinz, Nucl. Phys. {\bf A566} 205c (1993).
%
\bibitem{crss} J.~Cleymans, K.~Redlich, H.~Satz and E.~Suhonen,
Z. f. Physik {\bf C58}, 347 (1993).
%
\end{thebibliography}
